\documentclass[twocolumn, showpacs,amssymb,aps,prl]{revtex4}

\usepackage{graphicx}

\begin{document}
\title{Dislocation Dynamics in Rayleigh-B\'enard Convection}
\author{Th. Walter$^1$,  W. Pesch$^1$,
and E. Bodenschatz$^2$ }
\affiliation
{\small \it $^1$ Physikalisches Institut der Universit\"at Bayreuth, D-95440
Bayreuth \\
$^2$  Laboratory of Atomic and Solid State Physics, Cornell
University, Ithaca, NY 14853}
\date{\today}

\begin{abstract}

Theoretical results on the dynamics of dislocations in Rayleigh-B\'enard
convection are reported both for Swift-Hohenberg  models  and the Boussinesq
equations.  For intermediate Prandtl numbers the motion of dislocations is
found to be driven by the superposition of two independent contributions: (i)
the Peach-Koehler force derived from the change of a Lyapunov potential with
pattern wave number; (ii) a non-potential advection force on the
dislocation core by its self-generated mean flow. Their competition allows for
the first time to understand the experimentally observed bound dislocation
pairs. 
\end{abstract}

\maketitle

Striped patterns are ubiquitous in nature. They are found in physical,
chemical, and biological systems, which are driven away from equilibrium
\cite{croho}.  In general natural patterns are not perfect due to the presence
of defects, like grain boundaries (line defects) and topological point defects
(dislocations) \cite{NePaLe_93}. The nucleation, motion, and annihilation of
dislocations is essential for many pattern-selection processes, which are
initiated by modulational instabilities \cite{croho}. Dislocations govern  the
ordering kinetics of initially disordered patterns \cite{Ha_00} and sustain in
defect turbulent systems the perpetual 
reordering of the planforms \cite{DaBo_02}.
Thus much effort has been devoted to the study of dislocations in striped
patterns (see e.g. \cite{croho, SiZi_81, TeCr_86, PlEgBo_98}). Dislocations 
present  a simple  realization of  topological singularities  in a field
description of continuous extended systems \cite{pismen}.  To which extent 
their dynamics can be understood in terms of ''particles'' subject to
effective forces is a general important issue, that transcends pattern
formation. 

Here we study dislocations in Rayleigh-B\'enard convection (RBC), 
which in the past has proven itself as a paradigm  for  pattern forming systems  \cite{croho,BoPeAh_00}.
In RBC a horizontal layer of a simple
fluid is heated from below and cooled from above. For temperature differences
$\Delta  T$  above a critical value $\Delta T_c$, buoyancy driven convection
sets in typically in form of striped convection roll patterns with the
critical wavenumber $q =q_c$. The basic hydrodynamic equations
(Oberbeck-Boussinesq equations, OBE) are well established and the theoretical
analysis can be quantitatively compared with well controlled experiments
\cite{BoPeAh_00}. The communality of RBC with other bulk pattern-forming
systems like, for example, vibrating granular layers \cite{deBrBiSh_98}, gas
discharges \cite{AsMuAmPU_98}, and Taylor-Couette flow \cite{croho} is most
clearly expressed in universal model equations, like the Swift-Hohenberg (SH)
equation \cite{SwHo_77} and its generalizations \cite{Ma_83}.

In this letter we concentrate on the dislocation dynamics in the case of gas
convection, which over the past decade has attracted most interest
\cite{BoPeAh_00}.  Besides the appropriate SH-model, for the first time the
full OBE equations are analyzed. We found the motion of dislocations to be
driven by two independent ''forces''.  One is the well known Peach-Koehler
force, which describes the tendency of the system to develop towards a striped
pattern with an optimal average wavenumber $\bar q  \approx  q_c$
\cite{SiZi_81,TeCr_86}. Of particular importance is  our explicit
identification  of a  second non-potential force due  to the advection of the
dislocation core by its self-generated, roll-curvature driven mean flow. We
demonstrate that this force acts in general to remove dislocations from the
system as to reduce  $\bar{q}$. The two forces may balance each other at a
certain background wave number $ \bar{q} = q_D < q_c$,  such that the
dislocation becomes stationary. For patterns with $\bar{q} \gtrsim q_D $ the
competition of the two forces   can prevent the annihilation of two
dislocations with  opposite topological charge. This explains for the first
time the experimentally observed bound dislocation pairs
\cite{BoCadeBrEcHuLeAh_92}. Although the significance of the mean flow had
been expected for a long time in line with experiments \cite{Wh76}, its direct
identification in a previous theoretical analysis failed \cite{rem_SiZi}.

\begin{figure}[h]
\centering\includegraphics[width=1.8in]{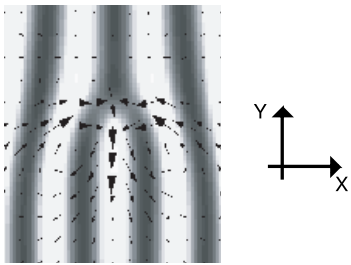}
\caption{Midplane temperature
field (gray scale) of a dislocation with positive 
topological charge $Q =2 \pi$ 
and the superimposed  mean-flow velocity field (arrows).  The
dislocation is moving downwards along    the $y$ -axis. From simulations
of the OBE  for $\epsilon \equiv  (\Delta T - \Delta T_c)/ \Delta T_c =0.3$,
$\sigma =1.2$  and background wavenumber ${\bar q}=  q_c$ (see text).
A dislocation with $Q = -2 \pi$ would move upwards in this case.
\label{fig:mean}} \end{figure}

In RBC the strength of the mean flow is determined by the Rayleigh number $R$
(a dimensionless measure of the temperature gradient across the convection
cell) and in particular by the Prandtl number $\sigma = \nu/\kappa$, with the
kinematic viscosity $\nu$ and the thermal diffusivity $\kappa$. Intermediate
Prandtl numbers  ($\sigma\sim 1$) are realized in gas convection
experiments \cite{BoPeAh_00}. In Fig. \ref{fig:mean},  a snapshot of the
midplane temperature field $\psi(x,y)$ containing a dislocation with positive
topological charge $Q =2 \pi$ from simulations of the OBE is shown.  $Q$  is
defined via the winding number of the phase gradient $\nabla \Phi(x,y)$ of
$\psi$  around a dislocation with the two possibilities $\oint \nabla
\Phi(x,y) ds = \pm 2\pi$. The mean flow (arrows),  which is driven by strong
curvature of the rolls in the vicinity of the dislocation core, is maximal
along the symmetry axis and  tends to advect the dislocation downwards out of
the system. In the direction perpendicular to the plane shown in Fig.
\ref{fig:mean} the mean flow shows an almost perfect  parabolic profile.

While a comprehensive exploration of the $ R, \sigma$ parameter space within the
OBE is extremely  time consuming, we  have gained much insight into the basic
physical mechanism of dislocation dynamics by first studying the standard
two-dimensional generalized  SH-equations  (GSH) \cite{Ma_83,TeCr_86,croho}:
\begin{eqnarray}\label{eq:swift}
\partial_t \psi +  (\vec{U} \cdot
\nabla)\psi &=& \left[\epsilon_S -
\left(1+\nabla^2 \right)^2\right]\psi-\psi^3,\\ 
\vec{U}=(U_x, U_y) &=&
(\partial_y \xi, -\partial_x \xi),\\ 
\label{wir} 
\left(\partial_t - \nabla^2 +c \right)\nabla^2 \xi &=& g
\left(\nabla \left( \nabla^2 \Psi \right) \times \nabla \psi
\right) \cdot \vec{e}_z.
\end{eqnarray}
In terms of the two-dimensional field $\psi({\vec x},t)$  the GSH-equations
describe (in suitable dimensionless units) the bifurcation to patterns with
critical wavenumber $q_c =1$ and their nonlinear saturation. $\epsilon_S
\propto \epsilon=  (\Delta T /\Delta T_c-1)$ measures the relative distance  to
onset.  The coupling to the mean drift  (or mean flow) $\vec U({\vec x})$ is characterized by
a  coupling constant $g$ (in RBC $g \propto 1/\sigma$)   and a cut-off parameter $c$. For
$g=0$ the GSH-equations derive from a Lyapunov functional $\cal L$
\cite{croho} and any dynamics is ``downhill"  towards the minimum of
$\cal L$.

In our numerical analysis we focused on dislocation climb, {\it i.e.}, the motion
parallel to the roll axis as shown in Fig. 1, in a  rectangular domain $
-L_{x,y}/2  < x, y < L_{x,y}/2$ with width $L_x$ and length  $L_y =2 L_x$. We
solved both the GSH- and the OBE equations numerically by a pseudo-spectral
method with semi-implicit time stepping \cite{DePe_94}. In order to minimize
finite-size effects we have simulated dislocation pairs with periodic boundary
conditions in the $x, y-$ directions (Fig. 1 has to be extended
mirror-symmetrically along the $y$-axis). We found, however, almost identical
results  for the less expensive simulations with a single dislocations  in a
box kept finite in the $y$ direction by gradually ramping $\epsilon$ to zero
at $y = \pm L_y/2$. The ideal pattern without an immersed  dislocation
consisted of up to $N = 64 $ roll pairs of a wavelength $ \lambda =
2\pi/q_{-}$, corresponding to $L_x = N \lambda$.  The typical numerical
resolution was $\ge 8$ grid points per roll diameter. It is crucial to study
systematically the defect dynamics as function of the  background wavenumber
$\bar q$  of the underlying pattern. A superimposed dislocation leads to $N
+1$ roll pairs in some parts of the cell (see Fig.1, below the dislocation)
and consequently to a reduced wavelength $\lambda_{+} = N/(N+ 1)  \ \lambda$ and
thus to a wavenumber $q_{+} = 2 \pi /\lambda_{+} > q_{-}$. We found that
defining the background wavenumber as $\bar{q} =(q_{+} + q_{-})/2$ was most
effective to absorb the finite-$N$ corrections. In designing the details of
our calculational scheme and to validate our SH-code we have considerably
profitted  from the comparison with Ref. \cite{TeCr_86}, devoted to the
SH-model.

First we discuss the solutions of the GSH-equations. The simulations (for
simplicity we consider the case of a single dislocation which climbs along the
y-axis) were  initialized with an approximate ansatz $\psi(x,y,
t=0) \propto cos[q_{-}\, x  - \phi(|\vec r|)]$ at  the center $ {\vec r_0} =
(x_0,y_0) = (0,0) $  of the simulation domain. Here $\phi(r)$ is the polar
angle about $ {\vec r_0}$.  After the initial transients  died out (say at $|y
-y_0| \gtrsim 4  \lambda $ in Fig.\ref{fig:mean}) the dislocation would climb
with a constant velocity $v_{st}$ and a relaxed asymptotic  shape $
\psi(x,y,t) = \psi(x, y- v_{st} t)$ was reached.  In subsequent runs,
starting from this state considerably reduced the computational time.  In
agreement with prior investigations \cite{TeCr_86,PlEgBo_98}, we have
identified a wavenumber $q_D(\epsilon, g)$, such that for $\bar{q} > q_D$ the
dislocation climbed downwards (the case shown in Fig. 1) with $v_{st} < 0 $
while it moved upward for $\bar{q} <  q_D$.

\begin{figure}[h]\includegraphics[width=3.5in]{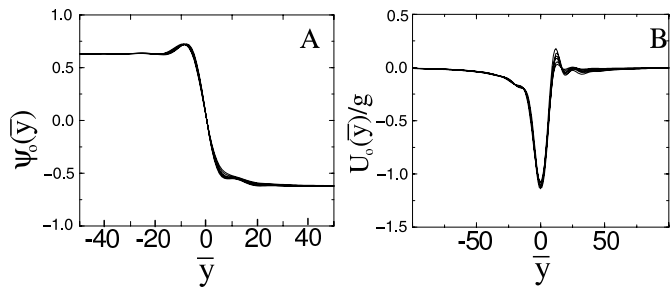}
\caption{Profiles
of  $ U_0(\bar y) \equiv  U_y
(x_0,\bar y)/g$ and $\psi_0(\bar y)\equiv  \psi(x_0,\bar y)$  plotted along the roll axis
centered about the dislocation core at $(x_0 =0, \bar y = y -v_{st} =0) 
$ obtained from  SH-simulations at $\epsilon_S = 0.3$  and $ {\bar q } =
0.98$ in the range $2 < g< 20$ for $N=16, q_{-} = 0.95$ and $c =2$.  
The profiles show  approximately  a
fifth  of the calculational grid of  $512$ pixels along $\bar y$, where    
 $16$ grid points   
correspond  to length of $ 3  \lambda  =  6 \pi /{q_{-}}  $. }
\label{fig:comb}\end{figure}

The key for the physical interpretation of  the results lies in a certain
universality of $\psi$ and $\vec U$. As demonstrated in Fig.~\ref{fig:comb}A
for a representative example of uniformly climbing dislocations   at {\it
fixed}  background wave number $\bar q = 0.98$ the numerical solutions
$\psi(x_0,  {\bar y}) \equiv \psi_0(\bar y)$ with ${\bar y} = (y-v_{st}\,t)$ of
Eq.~\ref{eq:swift} are virtually identical along the symmetry axis  ($x = x_0 =0$)
although the mean-field coupling $g$ varies considerably ($2<g<20$). In
addition, as shown in Fig.~\ref{fig:comb}B for the same range of parameters,
the (suitably reduced) mean flow component $U_y(x_0,  {\bar y}) /g$ along the
symmetry axis  has practically  a $g$-independent shape $ U_0(\bar y) $, while
the transverse component $U_x$ vanishes for symmetry reasons.
Inspection of Eq.(\ref{eq:swift}) suggests that $\psi(x_0,  {\bar y})$ can be
well approximated by the potential solution $\psi_0({\bar y})$ at $g =0$. This
applies also to the source term in Eq. (\ref{wir}) with respect to its
$\psi$-dependence.
Thus
Eq.\ref{eq:swift} on the symmetry axis reduces to:
\begin{equation}\label{eq:veloc_psi}
-v_{st}  \partial_{\bar y} \psi_0({\bar y}) = -\frac{ \delta {\cal L}}{\delta
\psi_0({\bar y})} + (U_y \partial_{\bar y} ) \psi_0({\bar y}). 
\end{equation}
Averaging Eq.(\ref{eq:veloc_psi}) over $\bar y$ 
leads to the following
relation for $v_{st}$ as function of $\bar{q}$ and $g$:
\begin{equation} \label{eq:veloc}
v_{st} (\bar{q},g) =  
v_{pot}(\bar{q}, g \rightarrow 0) + c_1(\bar{q}) \, g 
\, U^M_{0} \end{equation}
With respect to the advection term, we have made explicit  the (reduced) mean
flow $U^M_0$  at the dislocation core (i.e. the minimum of $U_0 (\bar y =0)$
in Fig.~\ref{fig:comb}B), which dominates the average in Eq.
(\ref{eq:veloc_psi}) also because of the strong maximum of $\partial_{\bar y}
\psi_0({\bar y})$  at $\bar y=0$ (see Fig. \ref{fig:comb} A) \cite{note_mean}.
The quantities $v_{pot}, c_1$  and $U^M_0$ have to be determined
numerically as function of $ {\bar q}$ and ${\epsilon_S}$.

For clarity the consequences of Eq. (\ref{eq:veloc})  are discussed for
$Q =2 \pi$  (see Fig. \ref{fig:mean}), since the case $Q = -2\pi$ is
analogous.   The potential or Peach-Koehler contribution
$v_{pot}$, originating from the functional derivative of $\cal L$ in Eq.
(\ref{eq:veloc_psi}),  is negative for $\bar{q} >  q_c = 1$, vanishes for
$\bar{q} \approx  q_c$, and becomes  positive for $\bar{q} < q_c$.   Thus the
dislocation climbs as to move  the wavenumber of the pattern in its wake
towards an  optimal value  $q_{opt} \approx q_c$, where the potential energy
has its minimum  \cite {SiZi_81} .  Since $c_1(\bar{q})$ in
Eq.(\ref{eq:veloc}) was found to be always positive  for all $ \bar q$ and
$\epsilon_S$, the mean flow ($\propto U^M_0 < 0$) would always  advect the
dislocation such as to decrease the wavenumber $\bar q$.  In fact, for 
$g \ne 0$ in patterns with $\bar q $  in the range $
q_D  \le {\bar q} \le q_{opt}$ the
potential contribution $v_{pot}$ 
is overcome by the mean flow
contribution leading to $v_{st}< 0$.   At ${\bar q} = q_D (\epsilon, g)$ both
effects cancel  leading to  a  immobile dislocation  ($v_{st}  = 0$).  In
agreement with prior investigations \cite{TeCr_86}  our numerical results show
that $v_{st}$ varies almost linearly as function $\bar{q}$ according to
$v_{st}  \propto  (\bar{q} - q_D)$ over a wide $\bar{q}$-range for fixed
$g>0 $ and $\epsilon_s > 0$.  According to Eq.(\ref{eq:veloc})  this implies at
fixed $\bar q$   a linear dependence of $v_{st}$ on $g$, which was
numerically confirmed   in a wide $g$-range ($ 0 \le g \le 30$).

\begin{figure}[h]
\includegraphics[width=3.5in]{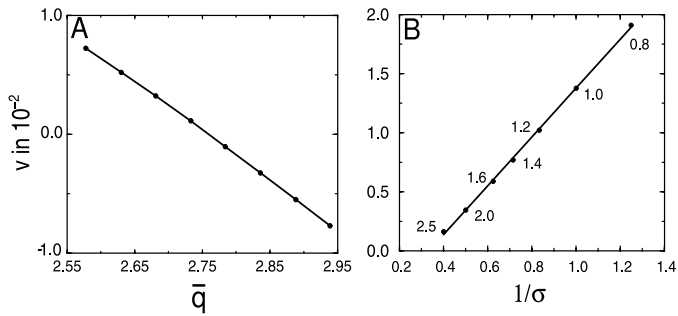}
\caption{Dislocation
velocity $ v_{st}$ in units of $10^{-2} d/t_v$ with $d$ the cell width, 
and $t_v = d^2/\kappa$ the vertical diffusion time.  A: as function of $\bar q$ at fixed $\sigma =
1.4$.  B:  as
function of $\sigma$ in the range $ 0.8 \le \sigma \le 2.5 $   of
at  $\bar q  = 2.91$,  data from simulations of the OBE for $\Delta T/ \Delta
T_c -1 = 0.3 $.} \label{fig:rb}
\end{figure}

Although in full hydrodynamics (OBE) a direct  analog  of potential $\cal
L$  in the GSH is not defined, the reasoning in terms  of a potential force (
$\sigma \rightarrow \infty$) and a competing mean flow force seems to apply well.  
As a typical example for $\sigma =1.4$,  we show in Fig. \ref{fig:rb}A  the
linear variation of $v_{st} (\bar  q)$ as function of $ \bar q$,
which crosses $v_{st}  =0$ at $\bar q = q_D = 2.75$.   In Fig.~\ref{fig:rb}B
it is demonstrated, that  $v_{st}$ varies like the SH-results linearly in
$\sigma ^{-1} \sim g$ at fixed $\bar q$.  Our findings are in perfect
agreement with the experimental and numerical analysis of the dynamics of the
off center ''giant spirals'' \cite{PlEgBo_98} for $\sigma \approx 1$ where the
uniformly rotating outer tip would probe roll patches in a fairly large $\bar
q-$ range.

In agreement with experiment  \cite{BoCadeBrEcHuLeAh_92} , from equation
(\ref{eq:veloc}) the possibility of bound dislocations pairs is obvious.
A well separated pair of oppositely charged  dislocations on a background
pattern with $q_D < \bar{q} < q_c$, is driven by the dominant mean-flow
contribution  $ \propto  |U^M_0|$ in Eq. (\ref{eq:veloc}) towards their
annihilation. However,  this advection force weakens continuously, as  the
opposing mean flow contributions from each dislocation begin to overlap
destructively. Eventually the dislocations stop moving, when the reduced
advection force is balanced by the opposing potential force. To our surprise
even the deceleration of the approaching dislocations was  well captured 
adiabatically by Eq. (\ref{eq:veloc}),  derived under the assumption  of 
constant $v_{st}$.  In fact, we found a direct  proportionality between the
decreasing velocity $v(t)$ of the approaching dislocations and the resulting $U^M_0
(t)$. The equilibrium distance $ l_D $ of the dislocations in the bound pair
as function of ${\bar q} $ and $g $, or $\sigma$, respectively, has not been
investigated systematically. However, typically we find $ l_D \sim 2 - 3
\lambda $ \cite{bound}.

In our OBE and GSH-simulations dislocations are found to be very robust
objects except when $\bar q$ lies in the vicinity of a cross-roll (CR)
instability boundary \cite{Bu_78} for ideal rolls, e.g. in the OBE near
$\bar q = 2.48 < q_c = 3.11$ for $\sigma = 1.4, \epsilon =0.3$. Then, as in
the experiments  \cite{AsSt_94,BoPeAh_00} the dislocation core might split
into a chain  of  bubbles by a localized transverse bridging of adjacent
rolls.    This local CR instability has been described at first in the
SH-model simulations \cite{NePa_92}, but  we doubt that the disclination
mechanism invoked there has a merit.

In conclusion, we have presented an analysis of the defect dynamics for
medium $\sigma$ and finite $R$ in RBC, which has allowed to separate clearly
potential- and mean-flow forces. Our main interest was to elucidate the basic
mechanism. Details, like for instance the dependence of $q_D$ on $\sigma$ and
$\epsilon$ and the impact of the stability properties of the background pattern will
given in a separate manuscript. We expect that our methodology, to
separate potential and nonpotential effects  will be applicable
as well  to a number of other  systems that show striped patterns as mentioned before. 
 E.B. is grateful for support from the National Science
Foundation under grant DMR0072077.


\begin{thebibliography}{999}
\bibitem{rem_SiZi} 
See the dicussion in \cite{ SiZi_81}, where the OBE  were solved,
however,   in a fairly small system with unrealistic free-slip boundary
condition.
\bibitem{note_mean} 
For numerical reasons it is easier to characterize the mean-flow
strength  at the dislocation core  by the average of the
profile $U_0(y)$  along $y$, which is  obviously proportional  to  $U^M_0$.
\bibitem{bound} 
In the OBE simulations we found  bound states for $\epsilon = 0.3, 
\sigma = 1.4$ in a regime $2.79 < q < 2.64$  slightly above $q_D = 2.75$ 
(calculation with $16$ rolls on a $256 \times  512$ grid).
\end{thebibliography}
\end{document}